\documentclass[12pt]{article}
\usepackage[cp1251]{inputenc}
\usepackage{amsfonts,amsmath}
\newtheorem{theorem}{Theorem}
\newtheorem{col}{Corollary}
\newtheorem{rema}{Remark}

\begin{document}

\large

\title {
On eigenfunctions of one-dimensional Schr\"odinger operator with
polynomial potentials
\author{Andrey E.~Mironov
\protect\footnote{ Sobolev Institute of Mathematics, Ac. Koptyug
avenue 4, 630090, Novosibirsk, Russia, and Department of Mathematics
and Mechanics, Novosibirsk State University, Pirogov street 2,
630090 Novosibirsk, Russia; e-mail: mironov@math.nsc.ru},
Bayan T. Saparbaeva \protect\footnote{Department of Mathematics and
Mechanics, Novosibirsk State University, Pirogov street 2, 630090
Novosibirsk, Russia; e-mail: saparbayevabt@gmail.com }}
\thanks{The work was supported by RSF (grant 14-11-00441).}}
\date{}
\maketitle

One-dimensional Schr\"odinger operator with polynomial poten\-tials of degree 3 and 4 appears in many areas of mathematical physics (see e.g. \cite{1}--\cite{3}).
In this paper we point out an connection between eigenfunctions of such operator and eigenfunctions of commuting ordinary differen\-tial operators
of rank two (rank two Baker--Akhiezer function).

We recall that if differential operators
$L_n=\sum_{i=0}^nu_{i}(x)\partial_x^{i}$ and
$L_m=\sum_{j=0}^mv_{j}(x)\partial_x^{j}$ commute, then by Burchnall--Chaundy  lemma \cite{4} there exist a non-zero polynomial
$R(z, w)$ such that $R(L_n, L_m)=0.$  The polynomial $R(z, w)$ defines the
{\it spectral curve} $\Gamma=\{(z, w)\in\mathbb{C}:R(z,
w)=0\}.$ If $\psi$ is an eigenfunction $L_n\psi=z\psi$, $L_m\psi=w\psi$, then $P=(z, w)\in\Gamma$. The {\it
rank} of the pair $L_n, L_m$ is $l=\dim\{\psi: L_n\psi=z\psi,
L_m\psi=w\psi\},$ where $(z, w)$ is a general point in $\Gamma$. In the case of operators of rank 1 eigenfunctions
are expressed via theta-function of the Jacobi variety of $\Gamma$ (see \cite{5}). At $l>1$ eigenfunctions correspond to the {\it spectral data}
(see \cite{6}) $S=\{\Gamma, q, k^{-1}, \gamma_1, ...,
\gamma_{lg}, \beta_1, ..., \beta_{lg}\},$ where $\Gamma$ is a Riemann surface of genus $g$, $q\in\Gamma$ is a marked point with a local parameter
$k^{-1}$, $\gamma_1, ...,
\gamma_{lg}\in\Gamma$ is a set of points and $\beta_j=(\beta_{j,1}, ...,
\beta_{j, l-1})$ is a set of vectors. {\it Baker--Akhiezer function of rank $l$} $\psi(x, P)=(\psi_1, ..., \psi_l),
P\in\Gamma$ is a function which satisfies the conditions:

\noindent{\bf 1}. On $\Gamma\setminus\{q\}$ function $\psi$ has poles in $\gamma_i$ and
          \begin{equation*}
          \operatorname{res}_{\gamma_i}\psi_j(x, P)=\beta_{i, j}\operatorname{res}_{\gamma_i}\psi_l(x, P).
          \end{equation*}

\noindent{\bf 2}. In the neighbourhood of $q$ the function $\psi(x, P)$ has the form
          \begin{equation*}
          \psi(x, P)=\bigg(\sum^{\infty}_{i=0}\xi_i(x)k_j^{-i}\bigg)\Psi(x, k),
          \end{equation*}
          where $\Psi(x, P)$ is a solution of the equation $\Psi_x=A\Psi$,
          \begin{equation*}
          A=\left(
          \begin{array}{cccccc}
           0 & 1 & 0 & ... & 0 & 0 \\
           0 & 0 & 1 & ... & 0 & 0 \\
           ... & ... & ... & ... & ... & ... \\
           0 & 0 & 0 & ... & 0 & 1 \\
           k+\omega_1(x) & \omega_2(x) & \omega_3(x) & ... & \omega_{l-1}(x) & 0 \\
          \end{array}
          \right),
          \end{equation*}
          $\omega_j(x)$ are functional parameters.
For the spectral data in general position there is a unique Baker--Akhiezer function.
Herewith for a meromorphic function $f(P)$ on  $\Gamma$ with the pole in $q$ of order $n$ there is a unique operator $L_f$ of order $l n$ such that
$L_f\psi=f(P)\psi$. For such meromorphic functions $f$ and $g$ operators $L_f$ and $L_g$ commute.
At $l>1$ Baker--Akhiezer function can not be found explicitly. Nevertheless the operators themselves can be found
in some cases.  The operators of ranks two and three in the case of elliptic spectral curves were found in \cite{6}, \cite{7}.
A new method of finding rank two operators corresponding to hyperelliptic spectral curves were suggested in \cite{8} (see also \cite{9}--\cite{12}).
In particular, in \cite{8} it is proved that
$L_4=(\partial_x^2+\alpha_3x^3+\alpha_2x^2+\alpha_1x+\alpha_0)^2+ \alpha_3g(g+1)x$ commutes with $L_{4g+2}$.
Operators $L_4$, $L_{4g+2}$ define commutative subalgebra in the first Weyl algebra. At $g=1$ these operators firstly appeared in \cite{13}.
It is turn out that there is a relation between eigenfunctions of $L_4$ at $g=2,4$ and functions from the kernel of
$L_2=\partial_x^2+\alpha_3x^3+\alpha_2x^2+\alpha_1x+\alpha_0.$
Let $\varphi$ be a solution of $L_2\varphi=0.$
\begin{theorem}
1. Let $g=2$, $z$ be a solution of the equation
        \begin{equation*}
          z^2+4\alpha_2 z+12\alpha_1\alpha_3=0.
        \end{equation*}
Then $L_4\psi=z \psi,$ where
        $\psi=p\varphi,\ p(x)=6\alpha_3x+z+4\alpha_2.$

\noindent  2. Let $g=4$, $z$ be a solution of the equation
        \begin{equation*}
         z^3+20\alpha_2 z^2+16(4\alpha_2^2+13\alpha_1\alpha_3)z+320\alpha_3(7\alpha_0\alpha_3+ 2\alpha_1\alpha_2)=0.
        \end{equation*}
        Then $L_4\psi=z \psi,$ where $\psi=p\varphi,$
$p(x)=280\alpha_3^2x^2+20\alpha_3(z+16\alpha_2)
x+z^2+20\alpha_2 z+64\alpha_2^2+168\alpha_1\alpha_3.$
\end{theorem}
Let
$
 l_4=p^{-1}L_4p,\  l_{4g+2}=p^{-1}L_{4g+2}p
$
(the function $p(x)$ is pointed out in Theorem 1).
\begin{col}
At $g=2, 4$ operators $l_4$, $l_{4g+2}$, $L_2$ form a commutative ring modulo $L_2$, i.e.
\begin{equation*}
[l_4, L_2]= B_1 L_2, \quad [l_{4g+2}, L_2]= B_2 L_2,
\end{equation*}
where $B_1, B_2$ are some operators.
\end{col}
Using \cite{8} one can show that for $\alpha_3^3-4\alpha_2\alpha_3\alpha_4+8\alpha_1\alpha_4^2=0$ the operator
\begin{equation*}
L_4^{a}=(\partial_x^2+\alpha_4x^4+\alpha_3x^3+\alpha_2x^2+\alpha_1x+\alpha_0)^2 +2g(g+1)x(\alpha_3+2\alpha_4 x)
\end{equation*}
commutes with an operator of order $4g+2$ (in a partial case it was observed by V. Oganesyan \cite{14}).
Let $\varphi$ be a solution of the equation $(\partial_x^2+\alpha_4x^4+\alpha_3x^3+\alpha_2x^2+\alpha_1x+\alpha_0)\varphi=0.$
\begin{theorem}
  1. Let  $g=1$, $z=\frac{\alpha_3^2}{\alpha_4}-4\alpha_2.$
        Then  $L_4^{a}\psi=z \psi,$
        where $\psi=p\varphi, \ p(x)=4 \alpha_4x+\alpha_3.$

\noindent 2. Let $g=2$, $z$ be a solution of the equation
        \begin{equation*}
         z^2-\bigg(\frac{3\alpha_3^2}{\alpha_4}- 16\alpha_2\bigg)z+24\alpha_1\alpha_3+192\alpha_0\alpha_4=0.
        \end{equation*}
        Then
        $L_4^{a}\psi=z \psi,$
        where $\psi=p\varphi,$
   $p(x)=24\alpha_4^2x^2+12\alpha_3\alpha_4 x- 3\alpha_3^2+\alpha_4(z+16 \alpha_2).$
\end{theorem}
The operator $L_4^{b}=(\partial_x^2+\alpha_1e^{x}+\alpha_0)^2+\alpha_1g(g+1)e^{x}$
commutes with an operator of order $4g+2$ \cite{12}.
Let us denote by $\varphi$ a solution of the equation
$\big(\partial_ x^2+\alpha_1e^{x}+\alpha_0+\frac{1}{4}(g+\varepsilon)^2\big)\varphi=0.$
\begin{theorem}
  1. Let  $\varepsilon=0$, then
        $
         L_4^{b}\psi=-\frac{1}{4}g^2(4\alpha_0+g^2)\psi,
        $
        where $\psi=p\varphi,$  $p(x)=e^{gx/2}.$

\noindent  2. Let $\varepsilon=1$, then
        $
         L_4^{b}\psi=-\frac{1}{4}(g+1)^2(4\alpha_0+(g+1)^2)\psi,
        $
        where $\psi=p\varphi,$  $p(x)=e^{-(g+1)x/2}.$
\end{theorem}
\begin{rema}
Solutions of the equation
$(\partial_x^2+\alpha_1e^{x}+\alpha_0)\varphi=0$ are expressed via Bessel function, namely the change of the variable
$x=\ln\big(\frac{y^2}{4\alpha_1}\big)$
reduces this equation to
$(y^2\partial_y^2+y\partial_y+(y^2+4\alpha_0))\varphi=0$.
\end{rema}

The authors are grateful to B.A. Dubrovin for valuable discussions.

\end{document}